\documentclass[referee]{aa}
\usepackage{graphics}

\begin{document}

\def\kms{km\thinspace s$^{-1}$}
\def\cm2{cm$^{-2}$}

  \thesaurus{11     
              (09.13.2;  
               11.09.1;  
               11.09.4;  
               13.19.3)} 

\title{A Search for CO in the Local Group Dwarf Irregular Galaxy WLM}

\author{C.L. Taylor\inst{1,2}
\and U. Klein \inst{3}}

\institute{Ruhr-Universit\"at Bochum, Astronomisches Institut,
Universit\"atsstra\ss{}e 150, 44780 Bochum, Germany
\and Five College Radio Astronomy Observatory, University of Massachusetts,
619B Lederle Research Tower, Amherst, MA, 01003, USA 
\and Universit\"at Bonn, Radioastronisches Institut, Auf dem H\"ugel 71,
53121 Bonn, Germany}

\offprints{C.L. Taylor}

\date{Received / Accepted }

\authorrunning{C.L. Taylor and U. Klein}

\titlerunning{A Search for CO in WLM}

\maketitle

\begin{abstract}

We present $^{12}$CO J = 1$\rightarrow$0 and J = 2$\rightarrow$1 
observations of the low metallicity (12 + log(O/H) = 7.74) Local Group 
dwarf irregular galaxy WLM made with the 15 m SEST and 14 m FCRAO 
telescopes.  Despite the presence a number of HII regions, we find no CO 
emission.  We obtain low upper limits on the integrated intensity (I$_{CO} 
\leq 0.18$ K \kms\thinspace for CO (1$\rightarrow$0)).  The non-detection 
is consistent with the result of Taylor, Kobulnicky \& Skillman 
(\cite{TKS}), that dwarf galaxies below a metallicity of $\sim$ 7.9 
are not detected in CO emission.  WLM shows that this trend continues for
low metallicity galaxies even as their metallicities approach 7.9. 
These results are consistent with the models of the metal poor ISM by Norman 
\& Spaans (\cite{NS}).  By comparing our CO data with observations of star 
formation in WLM, we find evidence for a high CO to H$_2$ conversion 
factor. 

\end{abstract}

\section{Introduction}

The WLM galaxy is a dwarf irregular that sits on the edge of the Local 
Group.  Its distance is 0.9 Mpc (Minniti \& Zulstra \cite{MZ}), close
enough for detailed studies of the stellar populations both from the
ground (Minniti \& Zulstra \cite{MZ}) and from HST (Dolphin \cite{D00}).  
As is typical for dwarf galaxies, the metallicity of WLM is low, 12 + 
log(O/H) = 7.74 (Skillman, Terlevich \& Melnick \cite{STM}), while there 
is a comparatively large amount of atomic hydrogen (M$_{HI}$ = 
4.7~$\times$~10$^{7}$ M$_{\sun}$; Huchtmeier, Seiradakis \& Materne 
\cite{HSM}).  Morris \& Lo (\cite{ML78}) observed a single position
at the center of the galaxy in the CO J = 1$\rightarrow$0 line, obtaining
a 3$\sigma$ upper limit of 0.25 K.

CO has been difficult to detect in dwarf galaxies, a fact which is
now most often attributed to their low metal content (e.g. Taylor,
Kobulnicky, \& Skillman \cite{TKS}).  Taylor et al. (\cite{TKS}) found
evidence for a dependency of the integrated CO J = 1$\rightarrow$0
line intensity, I$_{CO}$, upon metallicity, such that only dwarfs with
metallicities greater than $\sim$ 8 on the 12 + log (O/H) scale were
detected, even with extremely low upper limits obtained on the more
metal poor galaxies.   The metallicity of WLM places it between the
detected galaxies and the more metal poor ones, making it a logical
candidate for a new search for CO emission.  The tremendous improvements
in receiver technology since the 1970's allow for much lower upper 
limits than could be obtained by Morris \ Lo (\cite{ML78}) and
for observations of multiple positions within the galaxy.  Here we
report on our search for CO emission in the WLM dwarf irregular galaxy.

\section{Observations and Data Reduction}

\subsection{SEST}

WLM was observed using the 15~m Swedish ESO Submillimeter Telescope (SEST) 
at La Silla, Chile, between 1999 January 10 - 16.  Observations were carried 
out simultaneously in the CO J = 1$\rightarrow$0 line at 115 GHz and the 
J = 2$\rightarrow$1 line at 230 GHz.  The beamsize at 115 GHz is 45\arcsec, 
and at 230 GHz is 23\arcsec.  Using the Low Resolution Spectrometers, 
velocity resolutions of 1.8 \kms\thinspace and 0.9 \kms\thinspace were 
obtained at 115 GHz and 230 GHz, respectively.  The receivers were tuned to 
the systemic velocity of the galaxy, -116 \kms. The system temperatures 
varied from 330 to 470 K at 115 GHz and from 250 to 420 K at 230 GHz.  The 
pointing was checked at the beginning of each observing session by observing 
the nearby SiO maser source R Aqr.

Observations were made of seven positions within the galaxy, each in 
proximity to the groups of HII regions observed by Hodge \& Miller 
(\cite{HM}).  In other Local Group dwarf irregulars, such as IC~10 and 
NGC~6822, CO emission has been found in the vicinity of HII regions 
(e.g. Wilson \cite{W92}; Wilson \cite{W95}).

\subsection{FCRAO}

Observations were also obtained with the Five College Radio Astronomy
Observatory 14~m telescope in New Salem, Massachusetts from 1 - 3
December 1999, 30 December 1999 - 2 January 2000 and 10 - 14 February
2000.  The SEQUOIA focal plane array receiver, consisting of 16 pixels
operating in the 85 to 115 GHz band, was used to observe the CO 
J = 1$\rightarrow$0 line.  The beamsize of the telescope at 115 GHz is
44\arcsec, and the spacing between pixels in the array is 88\arcsec.
The Focal Plane Array Autocorrelation Spectrometers (FAAS) were used to 
obtain a bandwidth of 80 MHz with 313 kHz (0.8 \kms) channels.  Because
of the southern declination of WLM the system temperatures were 
higher in the FCRAO data than in the SEST data, ranging from $\sim$
600 to 1000 K.  The pointing was checked every 2 hours each session by 
observing the SiO maser sources R Cas and T Ceph.

The telescope was pointed so that the central 2 columns of 4 pixels 
in SEQUOIA fell within the optical galaxy.  The outer two columns, 
although outside most of the optical light from WLM, were well within 
the bounds of the HI distribution (Huchtmeier, Seiradakis \& Materne 
\cite{HSM}).  Pixel 11 in the array was aimed to point at one of the 
prominent groups of HII regions identified by Hodge \& Miller (\cite{HM}).
The telescope was operated in a staring mode, where the position was
{\it not} changed to fill in the gaps between the SEQUOIA pixels. 

The SEQUOIA observations are complementary with the SEST data because
they are not biased to observing only in the vicinity of HII regions.
While HII regions are an indication of recent massive star formation,
and hence likely places to find molecular gas, CO emission can be
found in areas with no corresponding H$\alpha$ emission (e.g. IC~10,
Becker \cite{B90}; NGC~4214, Walter et al. \cite{We00}).  Figure~1 shows 
the location of the SEST and FCRAO observations compared to an optical 
image of WLM.

\begin{figure}
\resizebox{\hsize}{!}{\includegraphics{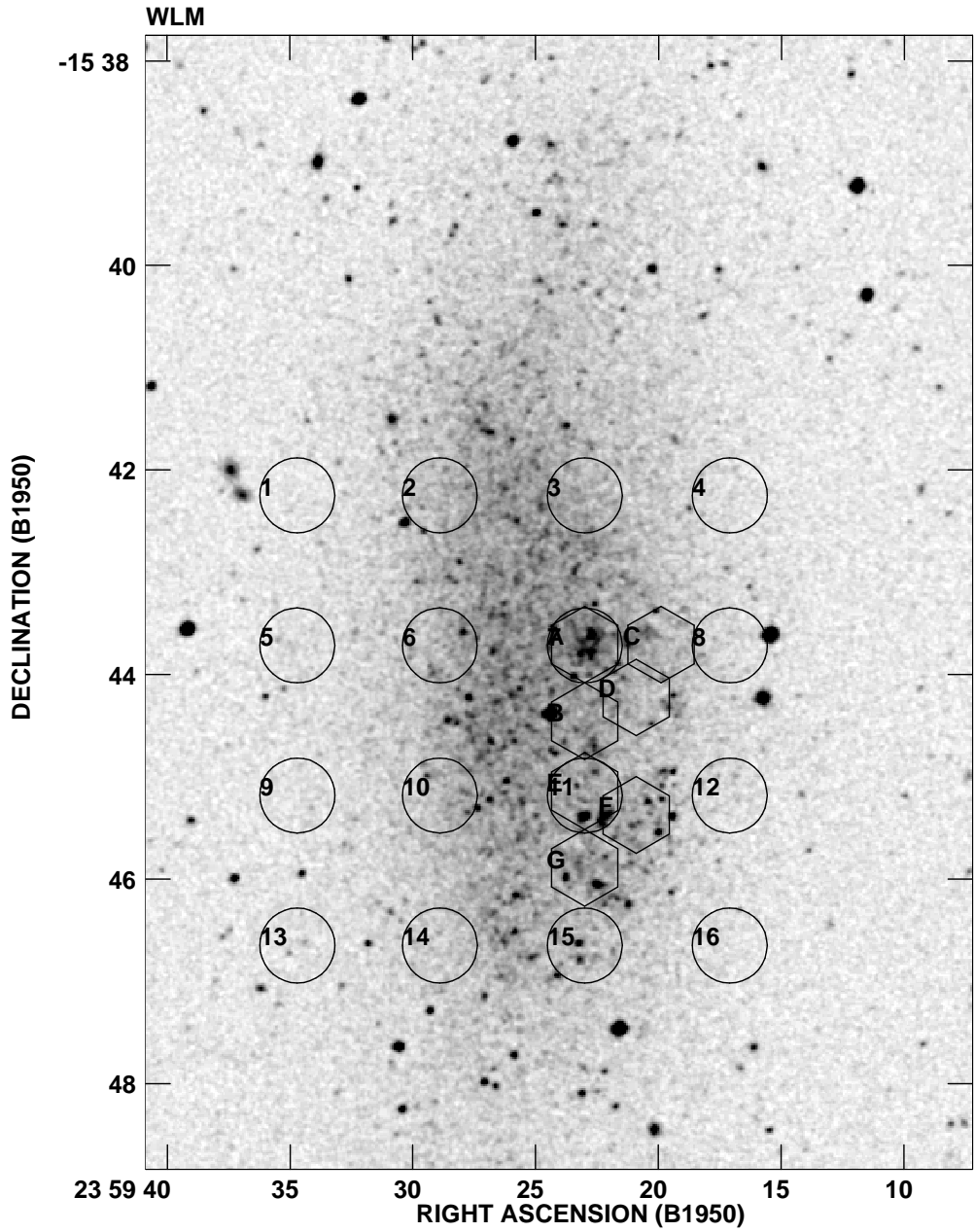}}
\caption{ Location of the SEST (hexagons) and FCRAO (circles) superposed
over a digitized POSS image of WLM.}
\label{fig1}
\end{figure}

\subsection{Data Reduction}

The data were reduced using the CLASS package.  We subtracted a linear
baseline from each individual scan; scans for which a linear baseline
was not sufficient were discarded.  The final spectra for each position
were obtained by averaging all the scans at each position.  To decrease
the rms noise in the final SEST spectra, we smoothed them to velocity
resolutions of 7.2 and 16.2 \kms\thinspace for the 1$\rightarrow$0 data 
and to 2.7 and 7.3 \kms\thinspace for the 2$\rightarrow$1 data.  To convert 
to the main beam temperature scale, main beam efficiencies ($\eta_{MB}$) of 
0.70 for 115 GHz and 0.5 for 230 GHz (SEST Handbook) were used.  The FCRAO 
spectra were smoothed to 7.2 \kms, and $\eta_{MB}$ = 0.45 was used.  All the 
analysis was done using the spectra at 7.2 \kms\thinspace resolution.
 
\section{Results}

Both the original resolution and the smoothed data were searched for
CO emission.  In neither the SEST data at either frequency, nor the
FCRAO 115 GHz data was any emission detected.
Table~1 gives the properties of the SEST data from the individual positions, 
including the offset in arcseconds from the center position, the rms noise 
in the spectra, and the 5$\sigma$ upper limit on integrated intensity 
(I$_{CO}$), for both the (1$\rightarrow$0) and (2$\rightarrow$1) data.
Table~2 gives the rms noise and (I$_{CO}$) 5$\sigma$ upper limits for the
FCRAO data.  In all cases the upper limits are calculated assuming 
undetected emission at the 5$\sigma$ level in one 7.2 \kms\thinspace 
channel.

\begin{figure}
\resizebox{\hsize}{!}{\includegraphics{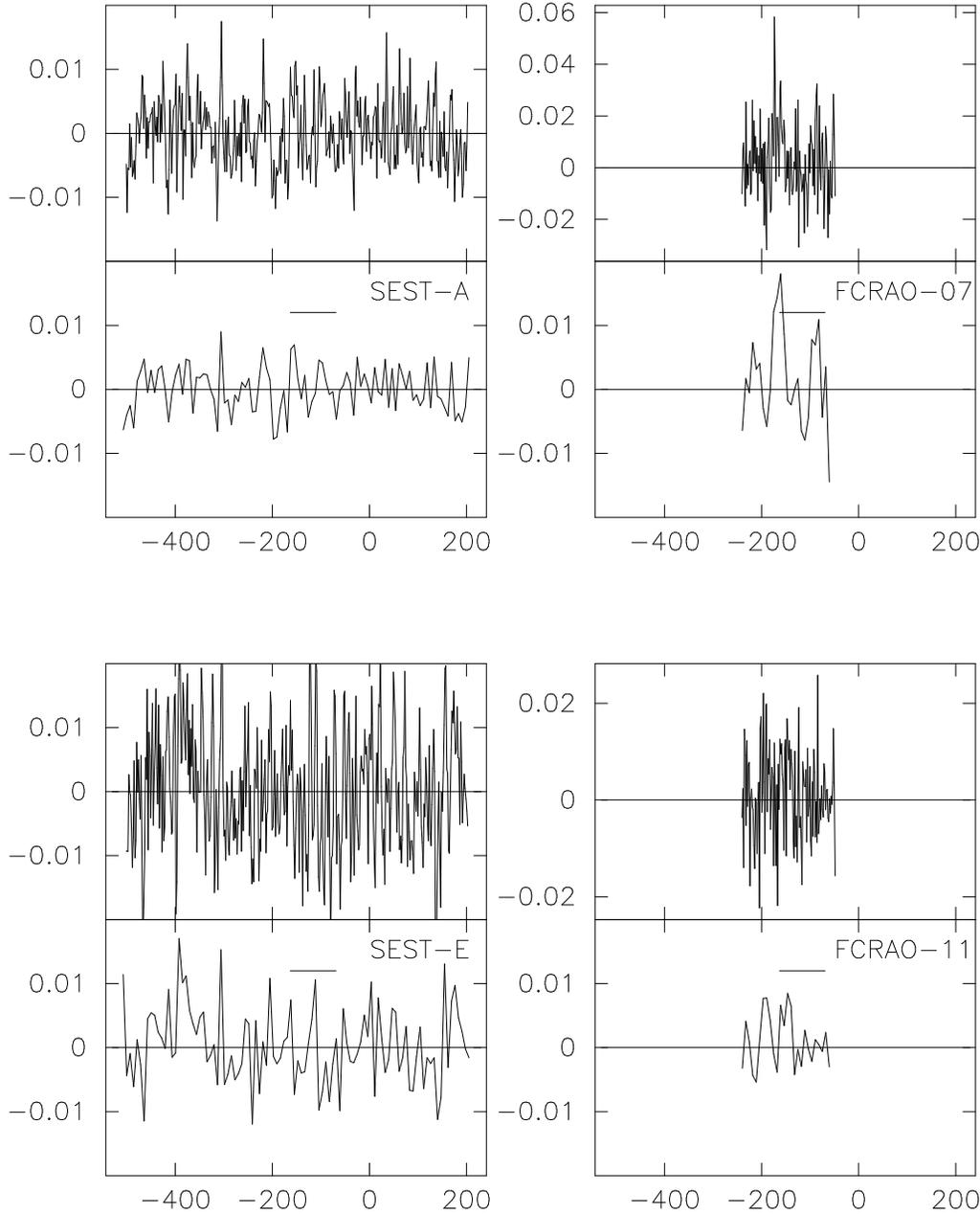}}
\caption{Example spectra from each telescope.  The upper spectra show the 
original velocity resolution, while the lower are smoothed to 7.2 \kms.  The 
horizontal line shows the full width at 20\% of maximum velocity range of 
the HI line from Huchtmeier et al. (\cite{HSM}).  SEST-A and FCRAO-07 are
the same location, as are SEST-E and FCRAO-11.}
\label{fig2}
\end{figure}

\begin{table}[t]
\begin{center}
{\bf Table~1}: SEST Positions in WLM
\end{center}
\begin{center}
\begin{tabular*}{14cm}{lcccccc}
\hline
Position & $\Delta$RA & $\Delta$Dec & $\sigma$(T$_{mb}$) (1-0) & I$_{CO}$ (1-0)  & $\sigma$(T$_{mb}$) (2-1) & I$_{CO}$ (2-1) \\
 & (arcsec) & (arcsec) & (K) & (K \kms) & (K) & (K \kms)  \\
\hline
A & 0 & 0 & 0.005 & $<$ 0.18 & 0.003 & $<$0.11 \\
B & 0 & -45 & 0.004 & $<$ 0.14  & 0.003 & $<$0.11 \\
C & -45 & 0 & 0.004 & $<$ 0.14 & 0.003 & $<$0.11 \\ 
D & -31 & -31 & 0.004 & $<$ 0.14 & 0.003 & $<$0.11 \\ 
E & 0 & -86 & 0.009 & $<$ 0.32 & 0.005 & $<$0.18 \\ 
F & -31 & -100 & 0.004 & $<$ 0.14 & 0.003 & $<$0.11 \\ 
G & 0 & -131 & 0.004 & $<$ 0.14 & 0.002 & $<$0.07 \\ 
\hline
\end{tabular*}
\vbox{ 
{\it Note}: The (0,0) position is 23$^h$ 59$^m$ 23.0$^s$, 
$-$15$^{\circ}$ 43$^{\prime}$ 42.5$^{\prime\prime}$.
}
\end{center}
\end{table}

\begin{table}[t]
\begin{center}
{\bf Table~2}: FCRAO Positions in WLM
\end{center}
\begin{center}
\begin{tabular}{lcclcc}
\hline
Position & $\sigma$(T$_{mb}$) (1-0) & I$_{CO}$ (1-0) & Position & $\sigma$(T$_{mb}$) (1-0) & I$_{CO}$ (1-0)  \\
 & (K) & (K \kms) &  & (K) & (K \kms) \\
\hline
1 & 0.011 & $<$ 0.40 & 9 & 0.016 & $<$ 0.58 \\
2 & 0.009 & $<$ 0.32 & 10 & 0.013 & $<$ 0.47 \\
3 & 0.011 & $<$ 0.40 & 11 & 0.009 & $<$ 0.32 \\
4 & 0.007 & $<$ 0.25 & 12 & 0.016 & $<$ 0.58 \\
5 & 0.011 & $<$ 0.40 & 13 & 0.011 & $<$ 0.40 \\
6 & 0.013 & $<$ 0.47 & 14 & 0.013 & $<$ 0.47 \\
7 & 0.018 & $<$ 0.65 & 15 & 0.009 & $<$ 0.32 \\
8 & 0.011 & $<$ 0.40 & 16 & 0.007 & $<$ 0.25  \\
\hline
\end{tabular}
\vbox{
{\it Note}: Position~11 corresponds to 23$^h$ 59$^m$ 23.0$^s$, 
$-$15$^{\circ}$ 45$^{\prime}$ 10.9$^{\prime\prime}$.
}
\end{center}
\end{table}

\section{Discussion}

To compare our WLM data with other gas rich dwarf galaxies, in Figure~3
we reproduce the plot of I$_{CO}$ versus galaxy metallicity from 
Taylor, Kobulnicky and Skillman (\cite{TKS}),  with a data point for
WLM added.  The data point we added represents position A from the
SEST observations.  This was chosen because its spectrum is typical,
being neither the lowest nor highest noise, and because this location
coincides with large HII regions, and so is a likely place to have
expected CO emission.  Taylor et al. noticed in their data a trend for
lower metallicity galaxies to be undetected while those of higher
metallicity were detected.  This was not a new result, but because
they used a relatively large number of galaxies, they found that the
division between detected and undetected galaxies occurred at 12 + log(O/H)
$\sim$ 7.9. (about 1/10 solar abundance).  WLM falls just below
this metallicity at 7.74 and in a region in the figure with a dearth
of data points.  Our non-detection of WLM at the level of I$_{CO} <$ 0.18 
K~\kms\thinspace in CO (1$\rightarrow$0) reinforces this result.  
Our new data point shows that the trend for galaxies with metallcities
below 7.9 to be undetected in CO emission continues almost right up
to a metallicity of 7.9.

Norman \& Spaans (\cite{NS}) have studied the formation of molecular
gas at high redshifts, where the metallicity is low.  They find that
the cold, dense phase of the ISM does not form until a metallicity of 
approximately 0.03 to 0.1 Z$_\odot$.  In their models, before this time
star formation occurs at very low levels.  At this metallicity C, O, and 
CO become abundant enough to contribute to cooling the gas, and giant
molecular clouds can form.  Because dust formation requires metals,
it is also in this metallicity range that there is sufficient dust
to shield molecular gas.  Also, the amount of molecular gas increases
because of the formation of H$_2$ on the surface of dust grains.
Although they made their models with protogalactic disks (presumably 
becoming spiral and elliptical galaxies with evolution), the phase
evolution of the ISM broadly fits what is known about dwarf galaxies.
Dwarf galaxies exhibit the lowest metallicities known from emission
line regions, down to Z $\leq$ 0.03 Z$_\odot$ (e.g. I Zw 18; Skillman
\& Kennicutt \cite{SK}).  The lowest metallicities seen in these galaxies
are close to the lower range of metallicity at which Norman \& Spaans argue 
that GMCs can form.  This would make sense if large star formation episodes
can only occur in gas cool enough and dense enough that it is molecular.
Our non-detection of WLM agrees with the Norman \& Spaans models as well.
The upper portion of the range 0.03 to 0.1 Z$_\odot$  corresponds to
12 + log(O/H) $\simeq$ 8.0, the metallicity at which dwarf galaxies are 
no longer detected in CO in Figure~3.  The metallicity of WLM, 7.74, is
below this level.

Taylor et al. (\cite{TKS}) attributed the non-detection of dwarfs at 
metallicities below 12 + log(O/H) $\simeq$ 8.0 to the CO to H$_2$ conversion 
factor (X$_{CO}$) having a dependence on metallicity which becomes sharply
non-linear at this metallicity. However, there are not enough data points in 
Figure~3 to say if there is truly a sharp cutoff in the detectability, or
if the small numbers make it appear so.  Clearly more observations, especially
of galaxies with metallicities in the range 7.8 $\leq$ 12 + log(O/H) $\leq$ 
8.0, are needed before the behavior of CO emission at low metallicities
can be understood.

\begin{figure}
\resizebox{\hsize}{!}{\includegraphics{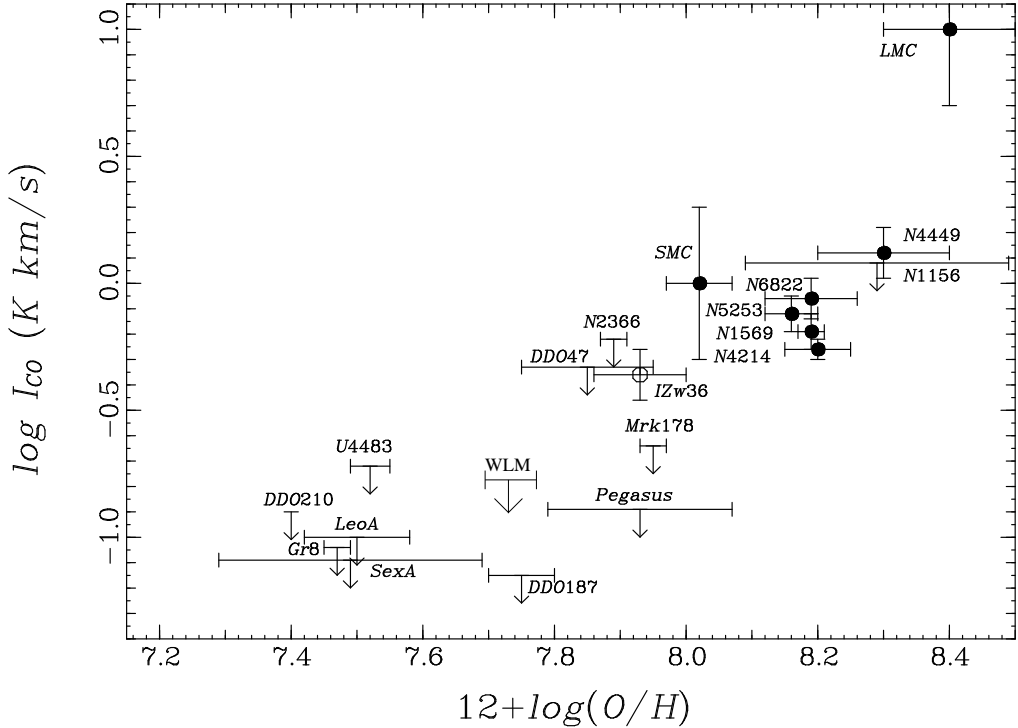}}
\caption{ Log(I$_{CO}$) vs. oxygen abundance, 12 + log (O/H).  The plot
is reproduced from Taylor, Kobulnicky \& Skillman (\cite{TKS}), with
the data point for WLM from this paper added.}
\label{fig3}
\end{figure}

Verter \& Hodge (\cite{VH}) obtained very low upper limits on CO 
emission in the nearby dwarf galaxy GR~8, and compared these limits to
the H$_2$ column density (N$_{H_2}$) inferred from the presence of 
star formation in the form of HII regions.  We will apply their analysis
to WLM, using the H$\alpha$ data of Hodge \& Miller (\cite{HM}):

\begin{equation}
\label{eq1}
\hfill  N_{H_2} = 9.5~\times~10^{-23}\ \frac{L_{H\alpha}(erg\thinspace s^{-1}) 
\tau_{SF}(yr)} {\epsilon_{SF} R^2_{mc}(pc)} \hfill
\end{equation}

where $L_{H\alpha}$ is the H$\alpha$ luminosity, $\tau_{SF}$ is the timescale 
of star formation, $\epsilon_{SF}$ is the star formation efficiency, 
and R$_{mc}$ is the radius of a molecular cloud. 
For $\tau_{SF}$ we adopt 10$^8$ yr, as did Verter \& Hodge.  This number 
ultimately derives from Lo, Sargent \& Young (\cite{LSY}), who argue
that $\tau_{SF}$ must be between the cloud-cloud collision time scale (the
timescale on which the gas will collapse to form stars) and the timescale
for disruption of the ISM by OB stars.   The former has an upper limit 
based upon the crossing time, which for dwarfs is $\sim$ 2~$\times~10^8$ yr, 
while latter will take place on a time scale of a few $\times~10^7$ yr.
Verter \& Hodge assumed $\epsilon_{SF}$ = 1 to obtain lower limits on 
N$_{H_2}$, but a more accurate value is 0.02 (e.g.  Taylor et al. \cite{THKG} 
for the post-starburst dwarf galaxy NGC~1569, Wilson \& Matthews \cite{WM}
for the giant HII regions NGC~604 and NGC~595 in M33). We will
estimate the column density with both efficiencies since this is the 
least secure property used in the calculation.   For R$_{mc}$
we adopt 10 pc based upon spatially resolved observations of giant molecular
clouds in nearby dwarf galaxies with interferometers (e.g. IC~10, Wilson
\& Reid \cite{WR} and Wilson \cite{W95}; NGC~6822, Wilson \cite{W94}).  

We can also estimate N$_{H_2}$ from our CO data, if we assume a CO to H$_2$
conversion factor.  For this we apply the conversion factor - metallicity
relation derived by Wilson (\cite{W95}) from Local Group galaxies, 
extrapolating down to the metallicity of WLM.  This gives a conversion
factor of 1.3~$\times~10^{21}$ \cm2 (K \kms)$^{-1}$, or 5.8 times greater 
than the Galactic value. Note that if the conversion factor behaves as 
suggested by Taylor et al. (\cite{TKS}), then the true conversion factor 
could be much higher than this estimate. Therefore, our N$_{H_2}$ estimates 
should be considered lower limits. 

In Table~3 we give the H$\alpha$ luminosity, the H$_2$ column density
estimated from star formation (N$_{H_2}$(SF)) for $\epsilon_{SF}$ = 0.02
and 1.0, and the H$_2$ column
density lower limit from the CO observations (N$_{H_2}$(CO)) for the
four SEST positions which overlap with HII regions from Hodge \& Miller 
(\cite{HM}).  For each position, the limits based on the CO data are 
already lower than the densities inferred from the star formation.  This
implies that the conversion factor is even higher than the extrapolation
of the Wilson (\cite{W95}) relation would suggest.  In order to get the
average N$_{H_2}$(CO) up to the level of the average N$_{H_2}$(SF) 
for the $\epsilon_{SF}$ = 0.02 case would require a conversion factor 
more than 1500 times the Galactic value.  Clearly this is extreme, and
may result from assuming too low a value for $\epsilon_{SF}$.  In
the case of $\epsilon_{SF}$ = 1 (which gives the minimum amount of
molecular material for the observed star formation rate), a conversion
factor of $\sim$ 30 times Galactic value would be necessary.  A more
realistic $\epsilon_{SF}$ would increase this number.  Through the
same analysis, Verter \& Hodge arrive at a lower limit for the for
the conversion factor in GR~8 of $\sim$ 30 times the Galactic value.
GR~8 has a lower metallicity than WLM (12 + log (O/H) = 7.5 compared to 
7.7), so in principle the conversion factor ought to be higher than in
WLM, but the fact that we obtain the same upper limit here as Verter \& Hodge
merely represents that both galaxies have similar levels of star formation 
and we have reached similar sensitivities in the observations.

Some dwarf galaxies have regions of CO emission with no corresponding
HII regions, areas in a pre-star formation phase (Becker \cite{B90}, Walter 
et al. \cite{We00}).  Such regions are not as common, however, as cases where 
the molecular gas and HII regions are in close association.  Clearly in the 
case of a molecular cloud with no star formation, one could not expect to 
estimate the molecular gas content associated with newly formed stars.  The 
crude calculation we describe above tells us nothing about such clouds.  But 
HII regions in the Galaxy are often associated with molecular gas on large 
scales, e.g. the Orion region.  This is also true of prominent HII regions
in dwarf galaxies like IC~10 (Becker \cite{B90}), NGC~1569 (Taylor et al.
\cite{THKG}) and NGC~4214 (Walter et al. \cite{We00}). It is unlikely that
all of the HII regions in WLM have totally disrupted their natal molecular 
gas, so it is reasonable to attempt to detect it, and to estimate the 
emission from these clouds based upon the young stars we currently see.

Of course there are large uncertainties in the above calculation: in 
addition to the star formation efficiency, one must assume an IMF (Verter 
\& Hodge use a Salpeter IMF),and neglect dust (observations of IC~10 have 
shown that dwarf irregulars can host considerable dust -- Bolatto et al. 
\cite{Be00}).   Substantial dust obscuration would mean we have 
underestimated the molecular gas implied by the observed star formation, 
which would further increase the conversion factor.  
Our data certainly do not prove a high CO-H$_2$ conversion factor,
but they are {\it consistent} with it.

\begin{table}[t]
\begin{center}
{\bf Table~3}: Star Formation Rates and Inferred H$_2$ Column Densities
\end{center}
\begin{center}
\begin{tabular}{lcccc}
\hline
Position & L(H$\alpha$) & N$_{H_2}$(SF) & N$_{H_2}$(SF) &  N$_{H_2}$(CO) \\
 &       & $\epsilon_{SF}$ = 0.02   & $\epsilon_{SF}$ = 1 &  \\
 &  (10$^{37}$ erg s$^{-1}$) & (10$^{21}$ \cm2)  & (10$^{21}$ \cm2) & (10$^{21}$ \cm2)\\
\hline
A & 2.0 & 95 & 1.9 & $>$ 0.24 \\
C & 1.0 & 48 & 0.95 & $>$ 0.18 \\
E & 0.8 & 38 & 0.76 & $>$ 0.42 \\
G & 2.1 & 100 & 2.0 & $>$ 0.18 \\
\hline
\end{tabular}
\end{center}
\end{table}

Klein (\cite{K00}) has recently argued that the conversion factor, 
X$_{CO}$,  depends upon the cosmic ray flux.   He used the 6 cm radio
continuum surface brightness as a tracer of cosmic rays and found that
as the cosmic ray content increased, X$_{CO}$ decreased.  The implication
here is that the heating by cosmic rays provides a significant portion
of the energy powering the CO emission, so that more CO emission per unit 
H$_2$ mass is observed when the cosmic ray flux is large compared to when 
it is small.  WLM fits into this picture well.  It has a 20 cm continuum
flux of only 5 mJy in the NVSS data (Condon et al. \cite{NVSS}), low compared
to other dwarf galaxies with measured 20 cm fluxes (e.g. H\"oppe et al. 
\cite{He94}, Klein, Weiland \& Brinks \cite{KWB}), suggesting a high X$_{CO}$.
The non-detection of CO is thus not unexpected.

\section{Summary and Conclusions}

We have searched for CO emission in the Local Group dwarf irregular galaxy 
WLM, and found none, obtaining low upper limits on the integrated 
intensity (I$_{CO} \leq 0.18$ K \kms\thinspace for CO (1$\rightarrow$0).  
WLM has many HII regions, indicating recent star formation activity, so the 
lack of CO emission may indicate a high CO to H$_2$ conversion factor, rather 
than a lack of molecular gas.  The non-detection is consistent with the result
of Taylor, Kobulnicky \& Skillman (\cite{TKS}) that dwarf galaxies with
metallicities less than 12 + log (O/H) $\simeq$ 8.0 are not detected in CO
emission while those with greater metallicities are.  This is explained
in the models by Norman \& Spaans (\cite{NS}), who argue that CO is not
abundant in the ISM until metallicities in the range $\sim$ 0.03 to 0.1
Z$_\odot$ (12 + log (O/H) $\sim$ 7.2 to 7.8).  Comparing our upper limits
to the molecular gas inferred from the presence of star formation in WLM
suggests that the CO to H$_2$ conversion factor may be as much as 30 times 
the Galactic value.

\begin{acknowledgements}
This work has been supported by the Deutsche Forschungsgemeinschaft
under the framework of the Graduiertenkolleg ``The Magellanic System,
Galaxy Interaction, and the Evolution of Dwarf Galaxies''.  The Five 
College Radio Astronomy Observatory is operated with the permission of 
the Metropolitan District Commission, Commonwealth of Massachusetts, and 
with the support of the National Science Foundation under grant AST-9725951.
\end{acknowledgements}

\end{document}